\documentclass[epjc3,twocolumn]{svjour3}

\RequirePackage[T1]{fontenc}

\smartqed  

\RequirePackage{flushend}
\RequirePackage[numbers,sort&compress]{natbib}
\RequirePackage[colorlinks,citecolor=blue,urlcolor=blue,linkcolor=blue]{hyperref}
\usepackage{graphicx}
\usepackage{dcolumn}
\usepackage{bm}
\usepackage{amsmath}
\usepackage{amssymb}
\usepackage{multirow}
\usepackage{enumerate}
\usepackage{amsfonts}
\usepackage{ifthen}
\usepackage{psfrag}
\usepackage{slashed}
\usepackage{hyperref}
\usepackage{gensymb}
\usepackage[utf8]{inputenc}
\usepackage{color}
\usepackage{subfigure}
\usepackage{url}

\newcommand{\be}{\begin{equation}}
\newcommand{\ee}{\end{equation}}
\newcommand{\bea}{\begin{eqnarray}}
\newcommand{\eea}{\end{eqnarray}}

\journalname{Eur. Phys. J. C}

\begin{document}

\title{Soft Scattering Evaporation of Dark Matter Subhalos by Inner Galactic Gases}

\author{Xiao-jun Bi\thanksref{addr1,addr2}
        \and
        Yu Gao\thanksref{e1,addr1}
        \and
        Mingjie Jin\thanksref{addr3}
        \and
        Yugen Lin\thanksref{e2,addr1,addr2}
        \and
        Qian-Fei Xiang\thanksref{addr4}
}

\thankstext{e1}{e-mail: gaoyu@ihep.ac.cn}
\thankstext{e2}{e-mail: linyugen@ihep.ac.cn}

\institute{Key Laboratory of Particle Astrophysics, Institute of High Energy Physics, Chinese Academy of Sciences, Beijing, 100049, China\label{addr1}
          \and
          School of Physical Sciences, University of Chinese Academy of Sciences, Beijing, 100049, China\label{addr2}
          \and
          Department of Physics, Beijing Normal University, Beijing, 100875, China\label{addr3}
          \and
          Center for High-Energy Physics, Peking University, Beijing, 100871, China\label{addr4}
}

\date{Received: date / Accepted: date}

\maketitle

\begin{abstract}
The large gap between a galactic dark matter subhalo's velocity and its own gravitational binding velocity creates the situation that small subhalos can be evaporated before dark matter thermalize with baryons due to the low binding velocity. In case dark matter acquires an electromagnetic dipole moment, the survival of low-mass subhalos requires stringent limits on the photon-mediated soft scattering. The current stringent direct detection limits indicate for a small dipole moment, which lets DM decouple early and allows small subhalos to form. We calculate the DM kinetic decoupling temperature in the Early Universe and evaluate the smallest protohalo mass. In the late Universe, low-mass subhalos can be evaporated via soft collision by ionized gas and accelerated cosmic rays. We calculate the subhalos evaporation rate and show that subhalos lighter than $10^{-5}M_{\odot}$ in the gaseous inner galactic region are subject to evaporation via dark matter's effective electric and magnetic dipole moments below current direct detection limits, which potentially affects the low-mass subhalos distribution in the galactic center.
\end{abstract}

\section{Introduction}
\label{sect:intro}

The Weakly Interacting Massive Particle (WIMP) is well-motivated for explaining the Universe's cold dark matter abundance and seeding structure growth~\cite{Scherrer:1985zt,Bertone:2004pz}. Below the weak scale, electrically neutral WIMPs can still acquire effective coupling to photons, e.g. via loop effects in case the WIMP couples to charged mediators. Such effective electromagnetic (EM) operators allow for efficient soft scattering between dark matter and in particular ionized/charged environmental particles.

For electrically neutral WIMPs, the leading effective operator is the dimension-5 EM dipole operator~\cite{Sigurdson:2004zp,Masso:2009mu}, as has been considered in DM annihilation~\cite{Fukushima:2013efa}, nucleus recoil~\cite{Banks:2010eh,Barger:2010gv,Pospelov:2000bq}, cosmic ray energy loss~\cite{Cappiello:2018hsu, Feng:2021hyz} and collider searches~\cite{Fortin:2011hv,Barger:2012pf}. Due to kinematic requirement or experimental thresholds, these searches often involve a significant amount of momentum transfer. At low momentum exchange, the collision between DM and charged particles is dipole-charge scattering. The cross-section has a well-known $|q|^{-2}$ divergence and $q$ is the transfer momentum. For the dipole-charge soft scattering, the momentum-weighted cross-section $\sigma_{T}$ is finite which characterizes the efficiency of soft momentum transfer. The integrated soft scattering is just as efficient as hard scattering in terms of energy exchange, and plays an important role in DM - plasma transport studies for captured dark matter inside stars~\cite{Freese:2008ur,Iocco:2008xb}.

In the Early Universe, if DM acquire a large EM dipole moment, soft scattering can delay the kinetic decoupling between DM and the SM sector, and affect overdensity growth by exchanging momentum with the ionized fraction of matter. Considering the current stringent direct dection limit on DM dipole moment~\cite{DarkSide:2018ppu,SENSEI:2020dpa,XENON:2019gfn}, DM can decouple from SM plasma early and form low-mass subhalos. Recent studies on subhalo formation with DM-baryon scattering~\cite{Tseliakhovich:2010bj,Erickcek:2015bda,Ali-Haimoud:2021lka,Tashiro:2014tsa,Gluscevic:2017ywp,Nguyen:2021cnb,Boddy:2018kfv,
Lambiase:2021xcj,Slatyer:2018aqg,Boddy:2018wzy,Xu:2018efh} and dwarf galaxies\cite{Wadekar:2019mpc}, including velocity-dependent scenarios~\cite{Mahdawi:2018euy,Barkana:2018lgd,Ooba:2019erm,Maamari:2020aqz,Buen-Abad:2021mvc}, yield non-trivial effects and corresponding constraints on DM-baryon interaction strength. Note the dipole-dipole scattering on neutral particles is not enhanced, and the impact on overdensity in the Universe's neutral phase will be suppressed by the small ionization fraction during the dark age.

At late time, ionized hotspots re-emerge in inner galactic regions, yielding a large ionized fraction in the form of heated gas and cosmic rays. For small subhalos located in such regions, their relative velocity to the host's gas $v$ is typically far greater than the subhalo's escaped velocity $v_{\rm esc}$ to escape the weakly bound subhalo. Therefore, dark matter particles can escape the subhalo by losing even a tiny fraction of its energy during the collision on galactic gas.

As $v \gg v_{\rm esc}$, this unbinding would only require a cross-section significantly lower than that for DM-gas thermalization, as only a small fraction of energy needs to be accumulatively transferred during the age of host galaxies. Our Milky Way is one such massive galaxy with its diffuse gas mostly ionized. We will calculate the soft dipole-scattering heating rate of DM by colliding with galactic hot gas and cosmic rays, and place an upper limit on the DM's dipole form factor by assuming the survival of subhalos in the ionized Galactic interior.

\begin{figure}[tp]
\begin{center}
\includegraphics[scale=0.5]{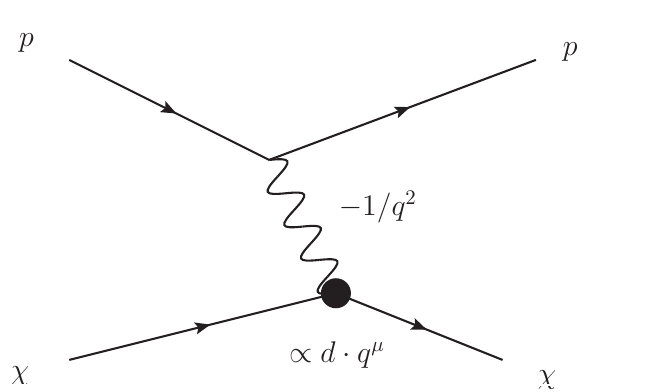}
\caption{The DM scatter with a charged particle (such as proton) and there is a typical $q^{-1}$ dependence in charge-dipole scattering amplitude.}
\label{fig:feynman}
\end{center}
\end{figure}

\section{Dipole-charge soft scattering}
\label{sect:model}

For low-energy collisions, we adopt the dark matter $\chi$ acquires effective EM dipole interaction~\cite{Barger:2010gv},
\be
\Delta {\cal L} = -\frac{i}{2} \bar{\chi}\sigma_{\mu\nu}(\mu+\gamma_5\mathcal{D})\chi F^{\mu\nu}
\ee
where the electric and magnetic dipole moments (EDM and MDM) $\mathcal{D}$ and $\mu$ derive from loop corrections of high-scale UV physics. Here we assume the DM acquire their abundance from some other high-scale interactions, thus these effective dipole moments can be small. Typical scenarios include $\chi$ being the neutral component of a gauge multiplet, or $\chi$ couples to the SM via heavier mediators, see Ref.~\cite{Hambye:2009pw,Kelso:2014qja,Ibarra:2015fqa} for related details. Non-zero dipole moments require the DM must not be self-conjugate. For fermion DM, $\chi$ needs to be either at least partially Dirac, or has multiple species; for simplicity we assume $\chi$ to be a Dirac fermion.

The scattering diagram of DM with a charged particle is shown in Fig.~\ref{fig:feynman}. The derivative on the effective interaction vertex picks up the photon momentum, which only partially cancels the photon propagator and leads to the well-known $q^{-2}$ dependence in the scattering cross-section. Notably dipole-charge scattering is the last infrared divergent diagram with EM operators. The higher-order EM anapole moment has $q^2$ on its $\bar{\chi}\chi\gamma$ vertex and would cancel the low-$q$ divergence and has no enhancement to soft scattering. DM self-scattering is dipole-dipole and is not soft-enhanced, which is different from self-interacting~\cite{Carlson:1992fn,Spergel:1999mh,Tulin:2017ara} and milli-charged~\cite{Li:2020wyl,Gabrielli:2015hua,Chun:2010ve,Melchiorri:2007sq} dark matter scenarios where heating via DM-DM scattering would become important~\cite{Bhattacharyya:2021vyd}.

We calculate the following two cases, corresponding to non-relativistic and relativistic scattering. For the non-relativistic collision between DM and ionized gas, their relative velocity is represented by $v$. The corresponding differential scattering cross-section is
\begin{equation}
\frac{{\rm d}\sigma}{{\rm d}\cos\theta}=\left\{\begin{array}{ll}
\alpha \mathcal{D}^{2}\frac{1}{v^{2}(1-\cos\theta)} & (\mathrm{EDM}) \\
\alpha \mu^{2} \frac{3 m_{\chi}^{2}+2 m_{\chi} m_{p}+2 m_{p}^{2}}{2\left(m_{\chi}+m_{p}\right)^{2}(1-\cos\theta)} & (\mathrm{MDM}).
\end{array}\right.
\label{eq:diff-cross}
\end{equation}
Due to that the cross-section of dipole-charge scattering has a divergence at low momentum exchange, so we use the transfer cross-section $\sigma_T$, which is finite and characterizes the efficiency of soft momentum transfer. So the corresponding transfer cross section is
\bea
{\sigma_T}\left(v\right)&\equiv &\int {\rm d}{\rm cos}\theta \frac{{\rm d}\sigma}{{\rm d} {\rm cos}\theta}(1-{\rm cos}\theta) \nonumber \\
&=&\left\{
                \begin{array}{ll}
2\alpha \mathcal{D}^2v^{-2} & (\rm{EDM})\\
\alpha \mu^2 \frac{ 3 m^2_\chi + 2 m_\chi m_p + 2 m^2_p}{(m_\chi + m_p)^2}  & (\rm{MDM}).
               \end{array}
\right.
\label{eq:simgabar}
\eea
There is a explicit $v^{-2}$ dependence in EDM induced non-relativistic collisions and the above results are the same as Ref.~\cite{Dvorkin:2013cea}. For the relativistic scattering between DM and cosmic ray proton, the corresponding elastic differential scattering cross-section is
\bea
\frac{{\rm d}\sigma}{{\rm d} T_\chi}=
\left\{
                \begin{array}{ll}
 \frac{e^2\mathcal{D}^2}{8\pi T_\chi |\textbf{p}|^2}(2 E^2 -2 E T_\chi-m_\chi T_\chi) &({\rm EDM}) \\
\frac{e^2\mu^2 }{8\pi T_\chi |\textbf{p}|^2}(2 |\textbf{p}|^2 -2 E T_\chi +m_\chi T_\chi) &({\rm MDM}).
                \end{array}
\right.
\eea
Considering that cosmic ray speed is close to the speed of light so DM can be seen as rest. $\textbf{p}$ represents 3-momentum of incident proton and $E$ is the proton's total energy i.e. $E=T_p+m_p$ where $T_p$ is proton's kinetic energy. $T_\chi$ denotes the kinetic energy of DM after collision and the momentum transfer to DM can be written as $|\textbf{q}|^2 = 2 m_\chi T_\chi$. The differential cross-section agrees with Ref.~\cite{Banks:2010eh} for a spin-$1/2$ collision target. Also, the corresponding transfer cross-section is
\be
{\sigma_T}=\left\{
                \begin{array}{ll}
\alpha \mathcal{D}^2 \left[1+m^2_p\left(\frac{1}{(m_\chi+m_p)^2+2m_\chi T_p}+\frac{2}{2m_p T_p +T^2_p}\right) \right] &({\rm EDM}) \\
\alpha \mu^2 \left[1+\frac{2m^2_\chi+m^2_p}{(m_\chi+m_p)^2+2m_\chi T_p} \right] &({\rm MDM}).
                \end{array}
\right.
\label{eq:simgabar2}
\ee

\begin{figure}[tp]
\begin{center}
\includegraphics[scale=0.6]{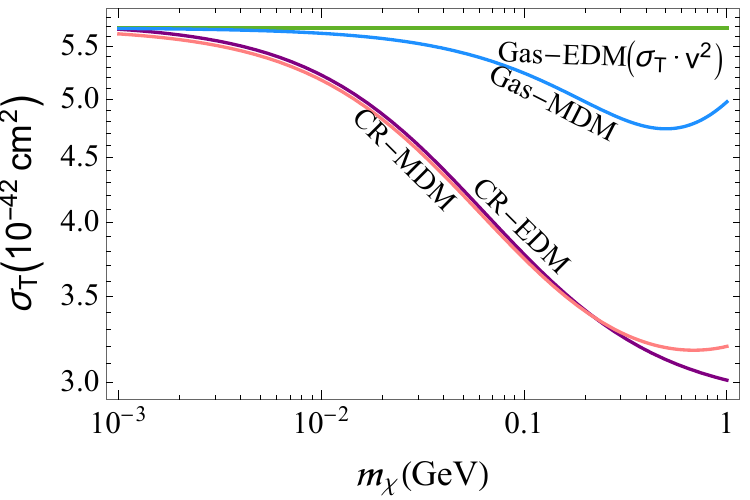}
\caption{Dark matter mass dependence of ${\sigma_T}$ in relativistic ($T_p=10$ GeV) and non-relativistic DM-proton collisions. The dipole moments ${\cal}D, \mu$ are chosen as $10^{-6}$ GeV$^{-1}$ which can satisfy the current direct detection limits. At the range of low dark matter mass, ${\sigma_T}$ becomes insensitive to $m_\chi$ and tends to be a constant.}
\label{fig:sigmabar}
\end{center}
\end{figure}

$\sigma_T$ is show-cased in Fig.~\ref{fig:sigmabar} where the transfer cross-section is plotted for relativistic collision with cosmic ray proton at 10 GeV, and non-relativistic collision with ionized gas. The dipole moments ${\cal}D, \mu$ are taken $10^{-6}$ GeV$^{-1}$ which can satisfy the current direct detection limits. Due to the $v^{-2}$ dependence in EDM induced non-relativistic collisions, $\sigma_T v^{2}$ is shown instead of $\sigma_T$. For MDM, the leading term is finite and not enhanced by $v^{-2}$. At the range of low dark matter mass, it is clear from Eq.~\ref{eq:simgabar} and Eq.~\ref{eq:simgabar2}, ${\sigma_T}$ becomes insensitive to DM mass and tends to be a constant.

It should be noted that there is a complication for heavy dark matter. For the relativistic case, when $m_\chi$ is much heavier than the proton mass, obviously collisions may not have sufficient energy to evaporate the DM in case $\frac{1}{2}m_\chi v^2_{\rm esc.}\gg T_p$. Therefore we will focus on the GeV and sub-GeV DM mass range, where the above equations remain valid and the heating effect is significant as well. Also, for the non-relativistic DM-gas collisions, The DM mass need to satisfy $m_\chi \ll m_p v^2/v_{\rm esc}^2$ to realize the prominent evaporation effects.

\section{Dark matter kinetic decoupling and the smallest protohalos}
\label{sect:decouple}

In this section we will calculate the DM's kinetic decoupling temperature and its temperature evolution, then we estimate the smallest dark matter protohalo size in our model. In the Early Universe, DM can keep the chemical equilibrium until DM annihilation rate becomes comparable to the Hubble expansion rate. After chemical decoupling, DM can still keep local thermal equilibrium by scattering with Standard Model(SM) particles in the thermal bath until the momentum exchange rate between DM and SM particles drops below the Hubble expansion rate. Later, DM completely decouple from the thermal bath and begin to stream freely without interacting with other particles, which is usually called the DM kinetic decoupling~\cite{Chen:2001jz,Visinelli:2015eka}.

\begin{figure}[tbp]
\begin{center}
\includegraphics[scale=0.5]{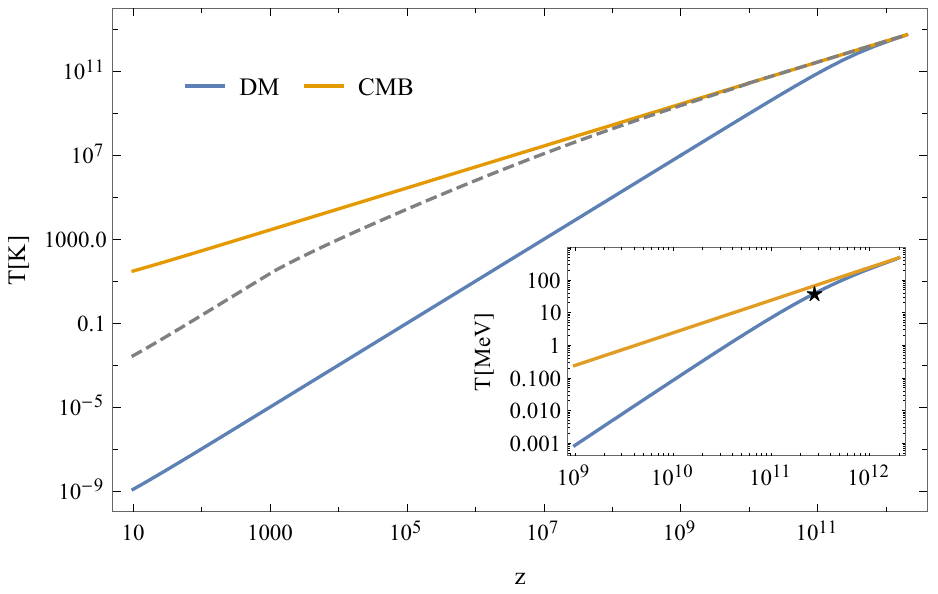}
\caption{The temperature evolution of DM and CMB with redshift. The CMB temperature is $T \propto a^{-1}$(yellow solid). In the early Universe, DM is coupled to the plasma $T_\chi=T \propto a^{-1}$. When $H(T_{kd})\thickapprox \gamma(T_{kd})$, the kinetic decoupling starts to occur. For a large dipole moment $10^{-3}$ GeV$^{-1}$ (gray dashed), the kinetic decoupling process can be slow. After recombination ($z\lesssim 10^3$), the cosmic ionization fraction decreases rapidly, which makes DM completely decouple. For a small dipole moment $10^{-6}$ GeV$^{-1}$(blue solid) within the current direct detection limit, the kinetic decoupling can occur quickly and the corresponding decoupling temperature is around 30 MeV which is marked with an asterisk symbol. After DM will decouple from the thermal bath and cold down as the expansion of the universe $T_\chi \propto a^{-2}$.}
\label{fig:temp_evo}
\end{center}
\end{figure}

The DM temperature evolution is related to the smallest protohalo size. To determine the
time evolution of DM temperature, we consider the Boltzmann equation for a flat FRW metric
\begin{equation}
E\left(\partial_t-H \mathbf{p} \cdot \nabla_{\mathbf{p}}\right) f=C[f].
\label{eq:Boltzmann}
\end{equation}
In the above equation, $(E,\mathbf{p})$ represents the energy and 3-momentum of DM and $f$ is the DM phase space density. $H=\dot a/a$ is the Hubble parameter and $a$ is the Universe scale factor. $C[f]$ is the collision term that describes the changes of $f$ between the scattering process of DM and relativistic SM particles. Following Ref.~\cite{Bringmann:2009vf}, it was the following form
\begin{equation}
C[f]=\gamma(T) m_\chi\left[m_\chi T \nabla_{\mathbf{p}}^2+\mathbf{p} \cdot \nabla_{\mathbf{p}}+3\right] f(\mathbf{p}).
\label{eq:collision}
\end{equation}
$\gamma(T)$ represents the momentum exchange rate  between DM and relativistic SM particles, which can be written as ($T$ is the plasma temperature)
\begin{equation}
\begin{aligned}
\gamma(T)= & \sum_i \frac{g_{\mathrm{SM}}}{6(2 \pi)^3 m_\chi^3 T} \int d k k^5 \omega^{-1} g^{ \pm}\left(1 \mp g^{ \pm}\right)\\
& \frac{1}{8 k^4} \int_{-4 k^2}^0 d t(-t) \overline{|\mathcal{M}|}^2.
\end{aligned}
\label{eq:momentum exchange rate}
\end{equation}
In the Eq.~\ref{eq:momentum exchange rate}, $(\omega,k)$ respectively represent relativistic SM particles' energy and momentum. Here we only consider the scattering between DM and charged SM particles through the dipole-charge interaction, so the sum is only taken over all possible charged SM scattering partners and $g_{\mathrm{SM}}$ represents the statistical degrees of freedom associated with charged particle species, spin and color and we refer the Ref.~\cite{Husdal:2016haj} for the evolution of the number of degrees of freedom. $g^{\pm}(\omega)=(e^{\omega/T}\pm 1)^{-1}$ is the distribution for Fermi or Bose statistics. $\overline{|\mathcal{M}|}^2$ represents invariant scattering amplitude squared for the process $\chi+i \rightarrow \chi+i$, which has been summed over final and averaged over initial spin states, and $t$ is the Mandelstam variable. When DM scatter with non-relativistic charged particles, the momentum transfer rate $\gamma$ can be expressed as
\begin{equation}
\gamma=\left\{\begin{array}{ll}
\frac{8 \alpha \mathcal{D}^{2} m_{\chi} \rho_{i}}{\sqrt{2\pi}\left(m_{i}+m_{\chi}\right)^{2}}(\frac{T_\chi}{m_\chi}+\frac{T_i}{m_i})^{-1/2}& \text { (EDM) } \\
\frac{12 \alpha \mu^{2} m_{\chi} \rho_{i}}{\sqrt{2\pi}\left(m_{i}+m_{\chi}\right)^{2}}\left[1-\frac{m_{i}\left(m_{i}+4 m_{\chi}\right)}{3\left(m_{i}+m_{\chi}\right)^{2}}\right](\frac{T_\chi}{m_\chi}+\frac{T_i}{m_i})^{1/2} & \text { (MDM) }.
\end{array}\right.
\label{eq:NR momentum exchange rate}
\end{equation}
For derivation details, see \ref{appendix:heating_rate}. After getting the momentum transfer rate $\gamma$, to find the evolution equation of DM temperature, multiplying Eq.~\ref{eq:Boltzmann} by $\mathbf{p}^2/E$ and integrating out $\mathbf{p}$, we can get
\begin{equation}
(1+z)\frac{d T_\chi}{d z}=2T_\chi+\frac{\gamma(T)}{H(z)}\left(T_\chi-T\right),
\label{eq:temperature evolution}
\end{equation}
and $T_\chi$ is defined by
\begin{equation}
\int \frac{d^3 p}{(2 \pi)^3} \mathbf{p}^2 f(\mathbf{p}) \equiv 3 m_\chi T_\chi n_\chi.
\end{equation}
From Eq.~\ref{eq:temperature evolution}, we can read off the two asymptotic behaviour of DM temperature: at high temperatures i.e. much greater than the kinetic decoupling temperature $T_{kd}$, DM is tightly coupled to the plasma and $T_\chi=T\propto a^{-1}$; at low temperature i.e. much lower $T_{kd}$, the DM temperature changes only because of the expansion of the universe and $T_\chi \propto a^{-2}$. The kinetic decoupling occurs when $H(T_{kd})\thickapprox \gamma(T_{kd})$. In generally, we need to numerically solve the Eq.~\ref{eq:temperature evolution} and the DM temperature evolution result is shown in Fig.~\ref{fig:temp_evo}.

In Fig.~\ref{fig:temp_evo}, the choice of parameter are $m_\chi=1$ GeV and dipole moment $10^{-6}$ GeV$^{-1}$ (blue solid) which can satisfy the current direct detection limits. Under the parameter, the kinetic decoupling temperature for EDM and MDM is comparable and the temperature evolution is similar so we only show the EDM case. The reason is that for scattering with relativistic particles, the transfer cross-section for EDM and MDM are almost same, as shown in Fig.~\ref{fig:sigmabar}, so that the corresponding kinetic decoupling temperature is almost consistent. From Fig.~\ref{fig:temp_evo}, we can obtain the kinetic decoupling temperature around 30 MeV which is marked with an asterisk symbol. Later DM will decouple from the thermal bath and the temperature will decrease as $T_\chi = T_{kd} (a_{kd}/a)^{2}$ due to the expansion of the universe.

We also show a situation with a larger dipole moment ${\cal}D$= $10^{-3}$ GeV$^{-1}$ (gray dashed) in  Fig.~\ref{fig:temp_evo}, where the decoupling process can be slow. When $H(T_{kd})\thickapprox \gamma(T_{kd})$, the kinetic decoupling starts to occur. However, if DM particles have a large electromagnetic coupling with the charged particles, DM will not decouple from the plasma quickly and the decoupling process will last for longer time. In the case, using $T_\chi = T_{kd} (a_{kd}/a)^{2}$ to describe DM temperature evolution is not accurate. After recombination ($z\lesssim 10^3$), the cosmic ionization fraction decreases rapidly, which makes DM completely decouple and its temperature will drop as $T_\chi \propto a^{-2}$. Considering the current stringent direct dection limit, for small dipole moments (blue solid), the kinetic decoupling can occur quickly and the kink at $z\approx 10^3$ are not obvious.

The kinetic decoupling temperature is closely related to the mass of the smallest dark matter protohalos. After $T_{kd}$, DM particles can stream freely from areas of high to low density without interacting with plasma so the process can erase the perturbations on scales smaller than the free-streaming length~\cite{Green:2003un,Green:2005fa}
\begin{equation}
\lambda_{fs}=a(t_0)\int_{t_{kd}}^{t_0} \frac{v(t)}{a(t)}d t.
\label{eq:fs}
\end{equation}
Above equation $a(t)$ is the Universe scale factor and $v(t)$ is the DM particle velocity which can be estimated as $v \approx \sqrt{T_{\chi}/m_\chi}$. The free-streaming length is the distance that DM can travel freely from the time of kinetic decoupling to present time $t_0$. The smallest protohalos from free-streaming effects can be estimated as the DM mass contained inside a sphere of radius $\lambda_{fs}/2$,
\begin{equation}
M_{fs}=\frac{4\pi}{3}\rho_m(t_0)(\frac{\lambda_{fs}}{2})^3 .
\label{eq:halo_mass}
\end{equation}
$\rho_m(t_0)$ is the dark matter density at the present time. In our model, for GeV scale dark matter and a kinetic decoupling temperature around 30 MeV, the corresponding smallest protohalo mass is around $10^{-7}M_\odot$.

Once we get the DM temperature evolution, as shown in Fig.~\ref{fig:temp_evo}, we can also calculate the corresponding Jeans scale that is a system's typical size for gravitational instability appearance and related to the DM temperature. When gravitational potential energy $U$ in a region surpass the thermal energy $K$ i.e. the total energy $U+K$ becomes negative, the jeans instability will occur which can lead to gravitational collapse. The critical case is corresponding to the Jeans scale~\cite{Jeans:1902fpv}
\begin{equation}
\lambda_J = c_s\sqrt{\frac{\pi}{G \rho_m}}.
\label{eq:jeans_scale}
\end{equation}
$c_s$ is the sound speed, for an ideal gas, $c_s \approx \sqrt{T/m}$. When $\lambda>\lambda_J$, the system will become unstable and gravitational perturbation can sustainingly grow. Its physical meaning is that once system scale is larger than Jeans length, sound pressures can't propagate the region in time to prevent the collapse. So the Jeans instability is actually the result of competition between thermal pressure and gravitational forces. The DM mass contained inside a sphere of radius $\lambda_{J}/2$ is the Jeans mass
\begin{equation}
M_J = \frac{4 \pi}{3} \rho_m\left(\frac{1}{2} \lambda_J\right)^3=\frac{\pi^{5 / 2}}{6} \frac{c_s^3}{G^{3 / 2} \rho_m^{1 / 2}}.
\label{eq:jeans_mass}
\end{equation}
In Fig.~\ref{fig:temp_evo}, DM temperature (blue solid) during structure formation (corresponding to $z\sim 20-30$) is around $10^{-8}$K. Using Eq.~\ref{eq:jeans_mass}, we can obtain the corresponding subhalo mass around $10^{-10}M_\odot$.

It should be emphasized that the Boltzmann equation and Jeans equation represent distinct physical processes. The former describes microscopic scattering of point particles, and the latter describes the long-range collective scattering between a particle and overdensities under gravitation. As a conservative choice, we adopt the larger of the two as the small-scale structure cut-off, i.e. the free-streaming scale to give the minimal halo mass. For dipole moment within current direct detection limits, halos with mass down to $10^{-7}M_\odot$ are allowed. Assuming such halos form, we would further study their evaporation in dense galactic areas at later times.

\section{Subhalo heating by galactic gases and cosmic ray}
\label{sect:model}

In the late Universe, for small subhalos located in inner galactic regions, they can be possibly evaporated by charged particles (mostly ionized gas and cosmic ray) via dipole-charge soft scattering. In this section, we will calculate the heating rate of subhalo by colliding with  galactic hot gas and cosmic ray. When DM collides with charged particles, the important physical quantity is the energy transfer rate. For the non-relativistic case of DM and ionized gas scattering, the DM's velocity is not negligible and the energy transfer rate needs to be averaged over the DM velocity distribution and that of the charged particle. For baryon gas, the thermally averaged energy transfer rate of per unit time is given by Ref.~\cite{Dvorkin:2013cea,Munoz:2015bca},
\begin{equation}
\begin{aligned}
\frac{\mathrm{d} \Delta E_p}{\mathrm{~d} t}= & \frac{m_\chi \rho_p}{\left(m_\chi+m_p\right)} \int d^3 v_p f_p\left(v_p\right) \int d^3 v_\chi f_\chi\left(v_\chi\right) \\
& \times \sigma_T\left(\left|\vec{v}_\chi-\vec{v}_p\right|\right)\left|\vec{v}_\chi-\vec{v}_p\right|\left[\vec{v}_{\mathrm{cm}} \cdot\left(\vec{v}_p-\vec{v}_\chi\right)\right]
\end{aligned}
\label{eq:thermal_xsec}
\end{equation}
where $\rho_p$ is gas density and $\sigma_T$ is the transfer cross section. $\vec{v}_{p}$ and $\vec{v}_{\chi}$ denote the `true' velocities of the proton and dark matter in the galactic frame, and $\vec{v}_{\rm cm}$ is the two-body center-of-mass velocity.

For thermal-averaged energy transfer rate, we assume DM velocity inside a subhalo follows a Maxwellian distribution,
\be
 f_{\chi}(\vec{v}_\chi)= \frac{1}{n}e^{-|\vec{v}_\chi-\vec{v}_0|^2/ \sigma^2_v}
 \label{eq:DM_velocity}
\ee
where $\vec{v}_0$ represents the subhalo's collective velocity that circulate around the galaxy. Using a Maxwellian distribution has been shown to be a good approximation which yields less than $\mathcal{O}(6 \%)$ level correction in energy transfer rate~\cite{Ali-Haimoud:2018dvo}. For protons in hot gas, their velocity follows a Boltzmann distribution that may also have a collective motion velocity $\vec{v}_{p0}$,
\be
 f_{p}(\vec{v_p})= \frac{1}{n}e^{-m_p|\vec{v}_p-\vec{v}_{p0}|^2/{2 k_B T}}
 \label{eq:proton_velocity}
\ee
DM subhalos at few kpc from the galactic center typically have $v_0\sim 10^{-4}$ and the relative velocity $v$ i.e. $|\vec{v}_{\chi}-\vec{v}_{p}|$ between the subhalo and gas is larger than the velocity dispersion inside the subhalo and gas. Namely, $|\vec{v}_0-\vec{v}_{p0}|$ dominates the relatively velocity $v$ in the DM-proton collision. The heating rate due to protons in gas is
\begin{equation}
\frac{{\rm d} \Delta E_{\chi}}{{\rm d} t}=\left\{\begin{array}{ll}
\frac{2 \alpha \mathcal{D}^{2} m_{p} m_{\chi} \rho_{p} v}{\left(m_{p}+m_{\chi}\right)^{2}} & \text { (EDM) } \\
3 \alpha \mu^{2}\left[1-\frac{m_{p}\left(m_{p}+4 m_{\chi}\right)}{3\left(m_{p}+m_{\chi}\right)^{2}}\right] \frac{m_{p} m_{\chi} \rho_{p} v^{3}}{\left(m_{p}+m_{\chi}\right)^{2}} & \text { (MDM) }
\end{array}\right.
\label{eq:heating_rate}
\end{equation}
where we have taken the limit where relative velocity $v$ is dominated by $|\vec{v}_0-\vec{v}_{p0}|$ and it is insensitive to the gas temperature. See \ref{appendix:heating_rate} for details. $\rho_p$ is the proton density and the additional heating by colliding with electrons can be obtained by replacing $m_p\rightarrow m_e$ and $\rho_p\rightarrow \rho_e=\rho_p\cdot m_e/m_p$.

For the relativistic case of DM and cosmic ray scattering, the heating rate is obtained by integrating transfer cross-section $\sigma_T$ in Eq.~\ref{eq:simgabar2} with the cosmic ray flux intensity $\Phi$
\bea
\frac{{\rm d} \Delta E_\chi}{{\rm d} t }&=& \int \Delta E_\chi  n v  ~ {\rm d} \sigma \nonumber \\
&=& \int {\rm d} T_i {\rm d} \Omega  \left(\frac{{\rm d} \Phi}{{\rm d} T_i {\rm d}\Omega}\right) \int \frac{{\rm d}\sigma}{{\rm d} T_\chi} T_\chi {\rm d} T_\chi
\label{eq:ene_ex}
\eea
where $n$ is the proton number density, $\Phi=nv$, $\Delta E_\chi =T_\chi$. The change of kinetic energy $\Delta E_k$ can be well-approximated as $\Delta  E_k\approx \frac{q^2}{2 m_\chi}$ for boosting $\chi$ from at rest, as an small momentum  exchange $q$ is transverse to the incident momentum. Larger $q$ is not perpendicular to the incident direction, yet the direction-dependence piece in $\Delta E_k$ is subleading if the transferred momentum dominates over the DM's initial momentum i.e. $q^2/2m_\chi \gg E_k^0$. Thus it applies to collisions where the DM temperature is negligible compared to the incident particle's energy.

Given this heating rate, the time scale for an average DM particle to be heated to its host subhalo's escaped velocity can be estimated as
\be
\tau_{\rm esc.} = \frac{1}{2} m_\chi\left( v_{\rm esc}^2- v_{\rm rms}^2\right) \cdot \left(\frac{{\rm d} \Delta E_{\chi}}{{\rm d}t} \right)^{-1},
\label{eq:tau_esc}
\ee
where $v_{\rm rms}$ is the root-mean-square velocity of DM inside the subhalo. Intuitively, neglecting thermal dispersion, one would expect Eq.~\ref{eq:tau_esc} be written in terms of $\sigma_T$. Simple comparison with Eq.~\ref{eq:simgabar} and Eq.~\ref{eq:heating_rate} yields,
\be
\tau_{\rm esc.} = \frac{(m_\chi+m_p)^2}{m_p^2}\cdot\left( \frac{v_{\rm esc}^2-v_{\rm rms}^2}{2v^2}\right) \left(
\sigma_T v n_p
\right)^{-1}.
\label{eq:survive}
\ee
Note the quantity in the last bracket is a thermalization time scale $\tau_{\rm th.}=(\sigma v n_p)^{-1}$, and the forefactor is highly suppressed due to $v_{\rm esc.}^2\ll v^2$, hence $\tau_{\rm esc}\ll \tau_{\rm th}$. Of course, our galaxy does not have enough baryons to kinetically thermalize dark matter, and such thermalization is never reached. This relation implies soft scattering is capable destabilizing small subhalos without DM and protons becoming thermally recoupled.

Stability of subhalos would require $\tau_{\rm esc.}> 10^{10}$ yr by collision with either gas or cosmic rays. Low-mass subhalos in inner galactic regions are most likely subjective to this evaporation effect, while large subhalos far away from the galactic center would be less affected.
DM particle's root-mean-square velocity $v_{\rm rms}$ and escaped velocity $v_{\rm esc}$ would depend on the subhalo size. We use an empirical scaling  relation for a particle's velocity dispersion $\delta_v$ from the Milky Way's observed subhalos~\cite{Walker:2009zp},
\be
\delta_v \approx  {3.9}{~\rm km/s}\left(\frac{M}{10^6 M_{\odot}}\right)^{1/3}
\label{eq:scaling}
\ee
to describe a population of sizable ($M \gtrsim 10^6M_\odot$) subhalos that contain visible stars. The scaling relation of velocity dispersion with the $1/3$ power-law of the halo mass is based on the predictions of the virial theorem. Generally lower mass subhalos are expected to exist in the galaxy. For our model, the possible small-scale structure has been elaborated in Sec.~\ref{sect:decouple}. In the $\rm \Lambda$CDM cosmology, small halos are seeded first then merge together to build-up large dark matter halos. During this hierarchical assembly, they have a self-similar virialized structure. Evrard et al.~\cite{Evrard:2007py}(included subsequent study~\cite{Lau:2009py,Munari:2013mh}) used a large set of dissipationless N-body simulations and showed that scaling relation for dark matter particles is fully consistent with theoretical predictions of the virial theorem. Besides, their study further demonstrated that the virial scaling relation displays a remarkable level of universality and self-similarity across a broad range of halo masses, redshifts, and cosmological models. So we can use this scaling relation at a lower mass range. The root-mean-square velocity $v_{\rm rms}$ and escaped velocity $v_{\rm esc}$ are related to the velocity dispersion $\delta_v$ by an ${\cal O}(1)$ factor, for a Maxwellian distribution in Eq.~\ref{eq:DM_velocity}, $v_{\rm rms} = 1.73 \delta_v$ and $v_{\rm esc}=2.44 \delta_v$.

\begin{figure}[tbp]
\begin{center}
\includegraphics[scale=0.6]{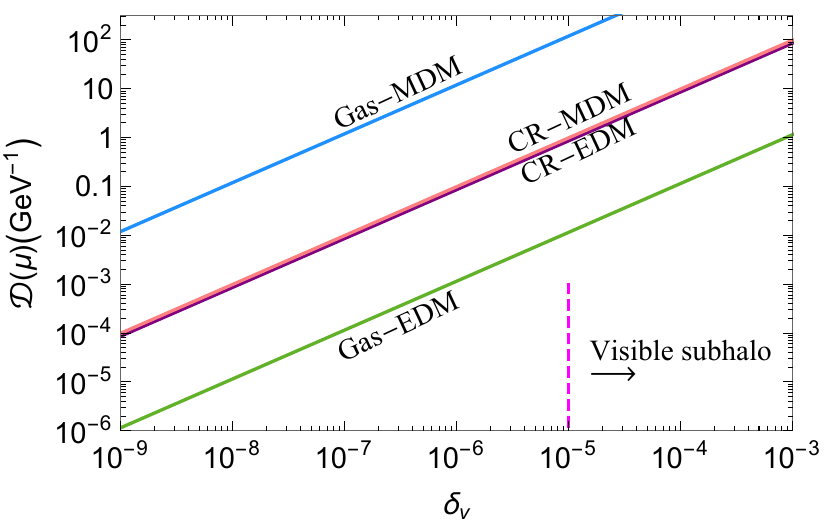}
\caption{Dipole moment $\mathcal{D}$ and $\mu$ limits for different subhalo's velocity
dispersion (normalized to the speed of light) that leads to $10^{10}$ yr evaporation in an environment with an average gaseous proton density $n_p=0.5$ cm$^{-3}$, or a cosmic ray flux intensity at twice of the locally measured intensity at the Sun. Subhalo velocity assumes $v=10^{-4}$ in the galactic frame and the velocity dispersion of visible subhalo ($M \gtrsim 10^6M_\odot$) is corresponding to $\delta_v\gtrsim10^{-5}$.}
\label{fig:bound_vb}
\end{center}
\end{figure}

\section{Galactic limits}
\label{sect:limit}

{\it Ionized gas} near the galactic disk heats up DM as subhalos travel through them. Galactic ionized clouds are categorized by temperature: (a) warm ionized medium (WIM) can distribute up to 50\% of the volume within 1 kpc from the disk, with 0.5 cm$^{-3}$ number density and 8000 K temperature. (b) hot ionized medium (HIM) at $10^{6-7}$ K temperature can distribute up to 70\% of volume within 3 kpc at a lower $10^{-2}$ cm$^{-3}$ number density~\cite{2018ApJ...862...34N}. The typical velocity of subhalos with ${\cal O}(kpc)$ orbit radius is $v \sim 10^{-4}$ 
and we use this velocity to estimate evaporation limits. For a larger radius or elongated orbit with a fraction of time $\epsilon$ in the gas-distributed region, the heating time scale should scale by $\tau_{\rm esc.}\rightarrow \epsilon^{-1}\tau_{\rm esc.}$, and the corresponding dipole moment limits scale by $\epsilon^{-1/2}$.

{\it Cosmic rays} are distributed in a cylindrical zone vertical to the galaxy disk, and the flux intensity $I_{\rm CR}$ depends on the distance to the cylinder axis and that to the galactic disk.  The cosmic ray proton energy spectrum is necessary to obtain the heating rate as in Eq.~\ref{eq:ene_ex}. So far the cosmic ray has only been measured `locally' at the Earth. But we mostly care about the region near the galactic center and the relative intensity distribution in other location can be modeled~\cite{Lipari:2018gzn} as,
\be
\frac{I(r,z)}{I(r_{\odot},0)}=\frac{{\rm sech}({r/r_{\rm CR})}}{{\rm sech}({r_{\odot}/r_{\rm CR})}}\cdot {\rm sech}({z/z_{\rm CR}})
\ee
where $I(r_{\odot}=8{\rm kpc},0)$ is the CR intensity at the sun, sech is the hyperbolic secant function, and $r,z$ denote cylindrical coordinates in kpc. The CR intensity steadily increases towards the center and decreases quickly off-disk. Modeling of the galactic cosmic ray flux varies in the diffusion zone height and size, yet as we are interested in the inner region, the distribution is relatively well described and is calibrated by measured values on the disk. With typical parameter choice $r_{\rm CR}=5.1$ kpc and $z_{\rm CR}\sim $ kpc, the volume-averaged proton flux within 1 kpc from the galactic center is about 2.1 times of that at the Sun's location. As an approximation, we will adopt the {\it shape} of the local energy spectrum~\cite{Amato:2017dbs} for cosmic ray protons, and use twice the measured magnitude as an estimate for the cosmic ray proton intensity in the inner galactic region. The energy spectrum is an approximate $E^{-2.7}$ powerlaw above the GeV scale. In the sub-GeV range, we use the spectrum recently measured by the Voyager satellite~\cite{Cummings:2016pdr}. Only proton flux is considered for cosmic rays, and the electron contribution is neglected due to its much lower flux intensity.

The dipole moment limits that lead to $\tau_{\rm esc.}= 10^{10}$ yr is shown in Fig.~\ref{fig:bound_vb} versus the subhalo’s velocity dispersion (normalized to the speed of light), with average proton density $n_p=0.5$ cm$^{-3}$ within 1 kpc where the CR flux intensity is 2 times of that at the solar system. In the figure, we fix the dark matter particle's mass $m_\chi=1$ GeV and using Eq.~\ref{eq:survive} to get the corresponding dipole moment limits for different subhalo’s velocity dispersion to lead to $10^{10}$ yr evaporation. Considering that cosmic ray travel close to the speed of light so the collisions are insensitive to subhalo's velocity and their limits on EDM and MDM are comparable. In non-relativistic gas-DM collision, the explicit $v^{-2}$ dependence in EDM $\sigma_T$ leads to faster heating than MDM and a significantly more stringent limit.

\medskip

Milky Way's observed subhalos are massive ($M \gtrsim 10^6M_\odot$) and typically have $\delta_v\gtrsim10^{-5}$ by using Eq.~\ref{eq:scaling}. One nearby example is the Canis Major substructure, $4.9\times 10^7$ $m_\odot$ in mass and 13 kpc from the galactic center~\cite{McConnachie:2012vd}. At such distance and mass the subhalo would be safe from evaporation as it spends a small fraction of time $\epsilon \ll 1$ in gaseous regions, and the required interaction strength ${\cal D} = \epsilon^{-1/2}\cdot 10^{-2}$ GeV$^{-1}$ is already excluded in direct detection searches.

\begin{figure}[tbp]
\begin{center}
\subfigure{\includegraphics[scale=0.6]{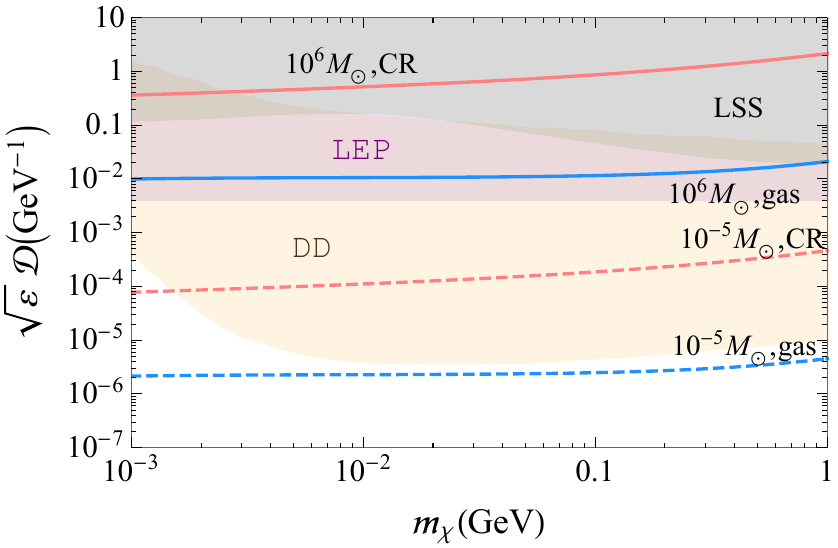}}
\subfigure{\includegraphics[scale=0.6]{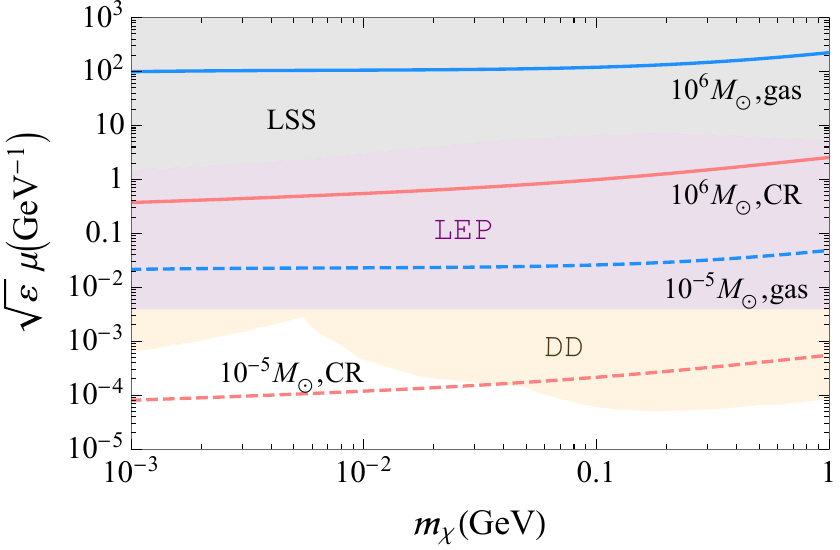}}
\caption{Dark matter EDM (upper) and MDM (lower) limits that leads to $10^{10}$ yr evaporation versus the different DM particle mass from soft non-relativistic collisional heating on gas(blue line) with the relative velocity $v=10^{-4}$ and relativistic scattering with cosmic rays(red line). Large-scale structures(LSS)~\cite{Chu:2018qrm}, collider (LEP)~\cite{Fortin:2011hv} and combined direct detection (DD) bounds~\cite{DarkSide:2018ppu,SENSEI:2020dpa,XENON:2019gfn} are also shown as color-shaped regions. The solid line represents the evaporation constraint of large mass visible halo ($10^6M_{\odot}$) and dashed line represents low mass invisible halo ($10^{-5}M_{\odot}$).}
\label{fig:limits}
\end{center}
\end{figure}

Lower mass, yet invisible DM subhalos are generally expected to form and become subject to evaporation effects. Again, using Eq.~\ref{eq:survive} and the scaling relation Eq.~\ref{eq:scaling}, we can get the corresponding dipole moment limits for specific subhalo size and different dark matter particle mass to lead to $10^{10}$ yr evaporation. As illustrated in Fig.~\ref{fig:limits}, the $\tau_{\rm esc.}=10^{10}$ yr limits from collisional heating on gas(blue line) and cosmic rays(red line) are given for subhalo mass at $10^6M_{\odot}$ (solid line) and $10^{-5}M_{\odot}$ (dashed line), which is respectively corresponding to the typical visible subhalo mass and a much lower invisible subhalo mass that allows the dipole-moment sensitivity dips below the current direct-search dipole limits. The temperature at which the DM kinetically decouples from the thermal bath must be at least 0.5 keV, in order to avoid the over-damping of large-scale structures(LSS)~\cite{Boehm:2004th,Bringmann:2006mu}. The corresponding constraint is shown as gray shaded region which is taken from Ref.~\cite{Chu:2018qrm}. Collider search constraint from LEP~\cite{Fortin:2011hv} is shown as purple region. The combined direct detection exclusion limits on DM dipole moment via DM-electron scattering, which includes DarkSide~\cite{DarkSide:2018ppu}, SENSEI~\cite{SENSEI:2020dpa} and XENON experiments~\cite{XENON:2019gfn}, are shown as orange shaped region. In sub-GeV mass range, comparing with these constraints, direct detection experiments give the strongest constraint on DM dipole moment.

In Fig.~\ref{fig:limits}, the evaporation limits given by the scattering with the ionized gas on ${\cal D}, \mu$ correspond to relative velocity $v=10^{-4}$ and average ionized gas density $n_p=0.5$ cm$^{-3}$. For a different relative velocity $v$, the EDM and MDM limits respectively scale by $v^{-1}$ and $v^{-3}$ due to the Eq.~\ref{eq:heating_rate} and~\ref{eq:tau_esc}. For satisfying the current sub-GeV direct detection exclusion limit on dipole moment strength,  the soft collisional heating can evaporate subhalo below $10^{-5}M_{\odot}$ in the inner galactic region and the corresponding escaped velocity is $v_{\rm esc}\approx10^{-8.5}$ (normalized to the speed of light) by using the scaling relation Eq.~\ref{eq:scaling}. Indirect limits based on DM annihilation are not shown as they can be model dependent.

In Sec.~\ref{sect:decouple}, we calculate the DM free-streaming scale to evaluate the smallest protohalo mass. Within current direct detection limits, subhalos with mass down to $10^{-7}M_\odot$ are allowed to form. Given the concentration of ionized baryons in the inner galactic region, for low mass subhalos located in a dense and heated galaxy inner region ($\sim kpc$), they may experience frequent collisions by dipole-charge interaction and will face an evaporation constraint. Considering the current direct-detection limits on DM dipole moment, Milky Way's hot ionized gas can evaporate subhalo below $10^{-5}M_{\odot}$ over a 10$^{10}$ yr time span which may affect the low-mass range of DM subhalo's mass distribution within $\sim kpc$ around the galactic center. In addition, soft scattering also causes cosmic rays to lose energy. Ref.~\cite{Cappiello:2018hsu} places a velocity-independent bound by examining spectral distortion in galactic cosmic rays. For dark matter particles with mass around MeV, their work gives the corresponding limit that the DM-proton cross-section around $10^{-25} \rm cm^2$, significantly above our limits (around $10^{-42} \rm cm^2$) for dipole moment $10^{-6}$ GeV$^{-1}$.

\section{Discussions}
\label{sect:summary}

To summarize, in this paper, for DM with an effective electromagnetic form factor, we have studied their kinetic decoupling in the Early Universe and calculate the corresponding smallest protohalo mass within current direct detection limits. In the late Universe, we investigated the evaporation of small subhalos by soft scattering between dark matter and ionized gas. The gap between a tiny gravitational binding velocity inside a small subhalo and the much larger relative velocity of subhalo to galactic gas renders the soft-collision evaporation very efficient. Satisfying the sub-GeV direct-detection limits, the inner Galaxy's hot ionized gas and cosmic rays are capable of evaporating low-mass subhalo below $10^{-5}M_{\odot}$ over a 10$^{10}$ yr time span. Evaporation by DM-gas collision potentially affects low-mass subhalo distribution around the galactic center, where DM is also abundant. Evaporation by cosmic rays would extend to a slightly larger region over a few kpc.

Visible subhalos of $10^{6}M_\odot$ are exempt from dipole-induced evaporation given existing direct detection limits, and subhalos far away from the galactic disk are also unaffected. Given the current direct detection limits, dipole-induced soft-scattering enhances the evaporation of $10^{-5}M_{\odot}$ or smaller subhalos, which are near the low-mass end of the galaxy's dark matter structure profile. In addition, low mass subhalos are also subject to other astrophysical effects, e.g. tidal disruptions. However, the cuspy central regions of subhalos can still remain relatively intact~\cite{Goerdt:2006hp}. Recently, Bosch et al.~\cite{vandenBosch:2017ynq,vandenBosch:2018tyt} used the high resolution N-body simulations to conclude that the complete physical disruption of subhalos is rare and tidal heating is not effective in the central regions of subhalos. In our work, the subhalos evaporation via dipole-charge scattering can be an additional contribution for the disruption of subhalos. Note our Milky Way is far from being an active galaxy. In galaxies with an active core, higher amount of fully ionized gas and stronger cosmic ray outflow would further enhance the evaporation due to soft scattering.

\medskip

{\bf Acknowledgments}

Authors thank Sujie Lin for communications about the galactic cosmic ray intensity distribution. Y.G. and Y.L. thank for support from the National Natural Science Foundation of China under Grant No.12150010 and the Institute of High Energy Physics, Chinese Academy of Sciences (E2545AU210). X.J.B is supported by the National Natural Science Foundation of China under Grant No.12175248.

\begin{appendix}
\section{Averaged heating rate}
In the appendix, we derive the energy transfer rate of non-relativistic case when DM particles scatter with charged particles (such as proton). The general expression is shown as Eq.~\ref{eq:thermal_xsec} in the main text and the velocity distribution of the dark matter and proton are shown as Eq.~\ref{eq:DM_velocity} and Eq.~\ref{eq:proton_velocity}. Generally, we can parameterize the transfer cross-section of the collision between dark matter and proton as $\sigma_T=\sigma_0 v^n$. Substituting the quantity into Eq.~\ref{eq:thermal_xsec} and integrating out the velocity distribution, the result can be expressed by two integrals~\cite{Munoz:2015bca}
\begin{equation}
\frac{{\rm d} \Delta E_{p}}{{\rm d} t}=\frac{m_{p} \rho_{\chi} \sigma_{0}}{\left(m_{\chi}+m_{p}\right)}\left[a I_{1}(n)+b I_{2}(n)\right]
\end{equation}
with $a=m_\chi/(m_\chi+m_p)$, $b=(T_\chi-T_p)/u_{th}^2(m_\chi+m_p)$. The $I_{1}(n)$ and $I_{2}(n)$ are respectively
\begin{equation}
\begin{aligned}
I_{1}(n)&=\frac{v^{n+6}}{(2 \pi)^{1 / 2} u_{\text {th }}^{3}} \int_{-\infty}^{\infty} d x e^{-x^{2} r^{2} / 2} \\
& \times x \frac{(x-1)(n+x+4)|x-1|^{n+3}}{(n+3)(n+5)} \nonumber ,
\end{aligned}
\end{equation}
\begin{equation}
\begin{aligned}
I_{2}(n)&=-\frac{v^{n+6}}{(2 \pi)^{1 / 2} u_{\text {th }}^{3}} \int_{-\infty}^{\infty} d x e^{-x^{2} r^{2} / 2} \\
& \times x \frac{(x-1)^{3}[(n+4) x+1]|x-1|^{n+1}}{(n+3)(n+5)}
\end{aligned}
\end{equation}
where $u_{th}$ is defined as $u_{th}^2=\frac{T_p}{m_p}+\frac{T_\chi}{m_\chi}$, $x=\frac{v_{th}}{v}$ and $r=\frac{v}{u_{th}}$.  $v$ represents the relative velocity between the subhalo and gas, and $v_{th}$ is the thermal relative velocity between them. First, we consider the energy transfer is dominated by their relative velocity $v$ and the thermal dispersion can be ignored. So we can take the corresponding limit $r\rightarrow \infty$. Substituting the expression of collisional cross-section between the gas and subhalo, as shown in Eq.~\ref{eq:simgabar}. For EDM, $n=-2$, $\sigma_0=2 \alpha \mathcal{D}^{2}$, the energy transfer rate is
\begin{equation}
\begin{aligned}
\frac{{\rm d} \Delta E_{p}}{{\rm d} t} & \approx \frac{m_{p} m_{\chi} \rho_{\chi} \sigma_{0} v} {\left(m_{p}+m_{\chi}\right)^{2}} \\
&=\frac{2 \alpha \mathcal{D}^{2} m_{p} m_{\chi} \rho_{\chi} v}{\left(m_{p}+m_{\chi}\right)^{2}}.
\end{aligned}
\end{equation}
For MDM, $n=0$, $\sigma_0=3 \alpha \mu^{2} \left[1-\frac{m_{p}\left(m_{p}+4 m_{\chi}\right)}{3\left(m_{p}+m_{\chi}\right)^{2}}\right]$, the energy
transfer rate is
\begin{equation}
\begin{aligned}
\frac{{\rm d} \Delta E_{p}}{{\rm d} t} & \approx \frac{m_{p} m_{\chi} \rho_{\chi} \sigma_{0} v^3}{\left(m_{p}+m_{\chi}\right)^{2}} \\
&=3 \alpha \mu^{2}\left[1-\frac{m_{p}\left(m_{p}+4 m_{\chi}\right)}{3\left(m_{p}+m_{\chi}\right)^{2}}\right] \frac{m_{p} m_{\chi} \rho_{\chi} v^{3}}{\left(m_{p}+m_{\chi}\right)^{2}}.
\end{aligned}
\end{equation}
The other case, in the Early Universe, the energy transfer is dominated by their relative difference of temperature and we can take the corresponding limit $r\rightarrow 0$. So we can get (for EDM)
\begin{equation}
\begin{aligned}
\frac{{\rm d} \Delta E_{p}}{{\rm d} t} & \approx \frac{4 m_{p} \rho_{\chi} \sigma_{0}} {\sqrt{2\pi}\left(m_{p}+m_{\chi}\right)^{2}u_{th}}(T_\chi-T_p) \\
&=\frac{8 \alpha \mathcal{D}^{2} m_{p} \rho_{\chi}}{\sqrt{2\pi}\left(m_{p}+m_{\chi}\right)^{2}}(\frac{T_\chi}{m_\chi}+\frac{T_p}{m_p})^{-1/2}(T_\chi-T_p).
\end{aligned}
\end{equation}
For MDM, the energy transfer rate is
\begin{equation}
\begin{aligned}
\frac{{\rm d} \Delta E_{p}}{{\rm d} t} & \approx \frac{4 m_{p} \rho_{\chi} \sigma_{0}u_{th}} {\sqrt{2\pi}\left(m_{p}+m_{\chi}\right)^{2}}(T_\chi-T_p) \\
&=\frac{12 \alpha \mu^{2} m_{p} \rho_{\chi}}{\sqrt{2\pi}\left(m_{p}+m_{\chi}\right)^{2}}\left[1-\frac{m_{p}\left(m_{p}+4 m_{\chi}\right)}{3\left(m_{p}+m_{\chi}\right)^{2}}\right] \\
& \cdot(\frac{T_\chi}{m_\chi}+\frac{T_p}{m_p})^{1/2}(T_\chi-T_p).
\end{aligned}
\end{equation}
The dark matter heating rate is obtained by substituting $p\leftrightarrow \chi$. Hence the heating rate of dark matter (dominated by their relative velocity i.e. $r\rightarrow \infty$) is
\begin{equation}
\frac{{\rm d} \Delta E_{\chi}}{{\rm d} t}=\left\{\begin{array}{ll}
\frac{2 \alpha \mathcal{D}^{2} m_{p} m_{\chi} \rho_{p} v}{\left(m_{p}+m_{\chi}\right)^{2}} & \text { (EDM) } \\
3 \alpha \mu^{2}\left[1-\frac{m_{p}\left(m_{p}+4 m_{\chi}\right)}{3\left(m_{p}+m_{\chi}\right)^{2}}\right] \frac{m_{p} m_{\chi} \rho_{p} v^{3}}{\left(m_{p}+m_{\chi}\right)^{2}}. & \text { (MDM) }.
\end{array}\right.
\end{equation}
and the energy transfer rate which is dominated by their relative difference of temperature (i.e. $r\rightarrow 0$) is
\begin{equation}
\frac{{\rm d} \Delta E_{\chi}}{{\rm d} t}=\left\{\begin{array}{ll}
\frac{8 \alpha \mathcal{D}^{2} m_{\chi} \rho_{p}}{\sqrt{2\pi}\left(m_{p}+m_{\chi}\right)^{2}}(\frac{T_\chi}{m_\chi}+\frac{T_p}{m_p})^{-1/2}(T_p-T_\chi)& \text { (EDM) } \\
\frac{12 \alpha \mu^{2} m_{\chi} \rho_{p}}{\sqrt{2\pi}\left(m_{p}+m_{\chi}\right)^{2}}\left[1-\frac{m_{p}\left(m_{p}+4 m_{\chi}\right)}{3\left(m_{p}+m_{\chi}\right)^{2}}\right](\frac{T_\chi}{m_\chi}+\frac{T_p}{m_p})^{1/2}(T_p-T_\chi) & \text { (MDM) }.
\end{array}\right.
\end{equation}
\label{appendix:heating_rate}
\end{appendix}

\bibliographystyle{spphys}
\bibliography{refs}

\end{document}